**Quantifying the influence of inter-county mobility patterns on the COVID-19 outbreak in the United States**

Qianqian Sun, Graduate Research Assistant
Maryland Transportation Institute, Department of Civil and Environmental Engineering
1173 Glenn Martin Hall, University of Maryland
College Park, MD 20742, United States, Email: qsun12@umd.edu

Yixuan Pan, Graduate Research Assistant
Maryland Transportation Institute, Department of Civil and Environmental Engineering
1173 Glenn Martin Hall, University of Maryland
College Park, MD 20742, United States, Email: ypan1003@umd.edu

Weiyi Zhou, Graduate Research Assistant
Maryland Transportation Institute, Department of Civil and Environmental Engineering
1173 Glenn Martin Hall, University of Maryland
College Park, MD 20742, United States, Email: wyzhou93@umd.edu

Chenfeng Xiong, Assistant Research Professor
Maryland Transportation Institute, Department of Civil and Environmental Engineering
1173 Glenn Martin Hall, University of Maryland
College Park, MD 20742, United States, Email: cxiong@umd.edu

Lei Zhang[*], Herbert Rabin Distinguished Professor (Corresponding Author)
Director, Maryland Transportation Institute
Department of Civil and Environmental Engineering
1173 Glenn Martin Hall, University of Maryland
College Park, MD 20742, United States, Email: lei@umd.edu

**Abstract**

As a highly infectious respiratory disease, COVID-19 has become a pandemic that threatens global health. Without an effective treatment, non-pharmaceutical interventions, such as travel restrictions, have been widely promoted to mitigate the outbreak. Current studies analyze mobility metrics such as travel distance; however, there is a lack of research on interzonal travel flow and its impact on the pandemic. Our study specifically focuses on the inter-county mobility pattern and its influence on the COVID-19 spread in the United States. To retrieve real-world mobility patterns, we utilize an integrated set of mobile device location data including over 100 million anonymous devices. We first investigate the nationwide temporal trend and spatial distribution of inter-county mobility. Then we zoom in on the epicenter of the U.S. outbreak, New York City, and evaluate the impacts of its outflow on other counties. Finally, we develop a "log-linear double-risk" model at the county level to quantify the influence of both "external risk" imported by inter-county mobility flows and the "internal risk" defined as the vulnerability of a county in terms of population with high-risk phenotypes. Our study enhances the situation awareness of inter-county mobility in the U.S. and can help improve non-pharmaceutical interventions for COVID-19.

**Background**

Since the World Health Organization declared the outbreak of COVID-19 as a Public Health Emergency of International Concern, the virus has created an increased threat worldwide. On March 13, 2020, the United States declared a national emergency, followed by various control measures at the state and county level. Due to the rapid human-to-human transmission of COVID-19, population flow is highly correlated to the spread of the disease[1-3]. Therefore, most control measures target restricting travel to mitigate the outbreak[1, 4-8]. Previous studies found that restrictions on long-distance travel are effective in the early stage while restrictions on local travel become more important after the virus disseminates[9-12]. However, there lacks research on interzonal population flow at a large scale, especially within the United States. In this study, we examine the inter-county mobility patterns under various control measures (declaration of a national emergency, stay-at-home order, partial reopening), which include both local and long-distance travel based on real-world observations. We also specifically investigate New York City, which saw the largest outbreak in the United States, from two aspects: its role in the inter-county mobility and the impact of outflow from NYC on the outbreak severity in those destination counties. Moreover, we quantify the impacts of inter-county travel on the spread of COVID-19 in the United States using a "log-linear double-risk" model. For each county, the model considers both the external risks (sum of inflow from infected areas weighted by the outbreak severity in origin counties), and the internal risks (the vulnerability based on high-risk phenotypes[13-15] including people age 65 and over, male, African Americans, and low-income households). Mobile device location data is known for its capability of capturing timely and real-world trajectories at a large scale. The population flow estimates from mobile device location data are believed valuable to help curb the pandemic[16-18]. We thus consider mobile device location data as an appropriate data source to estimate the inter-county mobility patterns. To retrieve the real-world and real-time mobility patterns, we utilize an integrated set of mobile device location data, which involves over 100 million anonymous devices at a monthly basis. Based on our previously developed algorithms[19], we have estimated daily inter-county trip tables between 3,143 counties and county-equivalents from January 1, 2020 to May 15, 2020 (weekends and holidays are removed from analysis).

**Nationwide spatiotemporal patterns of inter-county trips**

We first presented the temporal dynamics of nationwide inter-county trips in Fig 1a, where the total number of inflow inter-county trips of all the counties is examined. Based on the national trend, we recognized four stages: pre-pandemic (from January 1 to March 13), behavior change (from March 14 to April 13), quarantine fatigue (from April 14 to April 23), and partial reopening (after April 23). The pre-pandemic stage experienced a stable level before February 14 and a slight upward trend between February 15 and March 13, when the national emergency was declared. According to monthly variations of traffic volume, there was supposed to be an upward trend from January to March[20]. During the behavior change stage, the inter-county trip volume rapidly decreased within three weeks and then bottomed out on April 13, when the total inter-county trips decreased by 35%. After that, the inter-county trip total began to bounce back during the quarantine fatigue stage, which accompanied a consistent increase of confirmed COVID-19 cases. We defined this stage as quarantine fatigue because of an obvious rebound despite ongoing nationwide travel restrictions. On the week of April 24, states began deploying local phase-by-phase reopening plans by partially reopening selected businesses. During this stage, the inter-county trip volume kept rising even though the increase in confirmed cases had not slowed down.

We further evaluated the spatial differences between counties regarding the changes of total in-flow inter-county trips and such trips specifically from NYC (Fig. 1b, c, d, e). We calculated the weekday average from January 2 to January 31 excluding holidays as the baseline (Fig. 1b). Then we computed the weekday average week after week and the percentage change compared with the baseline. The week of March 9 presented widely increase instead of reduction (Fig. 1c) while the week of April 6 showed the most reduction (Fig. 1d). Overall, eastern counties present a more significant reduction than western counties. For the week of the national emergency declaration, 83% of counties still showed a percentage increase in the in-flow inter-county trips while the five counties in NYC show a slight reduction of inflow trips ranging from -5% to 0 (Fig. 1c). It may imply that travelers were cancelling trips to NYC or relocating themselves to places with lower infection risk. Compared with the daily average number of destination counties in January, 554 (17% of all counties), the inter-county trips from NYC still had a widespread distribution ending at 518 destination counties (16% of all counties). During the week of April 6, 92% of counties experienced a percentage reduction in in-flow inter-county trips, and the spatial distribution of inter-county trips from NYC was clearly narrowed down to 272 (8%) destination counties. During the most recent week, 48% of counties show a percentage increase in inflow trips, especially in western counties (Fig. 1e).

After evaluating the inflow inter-county volume at each county, we specifically investigated the greater New York City (NYC) area, consisting of New York County, Bronx County, Queens County, Kings County, and Richmond County, due to its intensive travel interactions with other counties. As the largest transportation hub and the most densely populated city in the United States, NYC became the U.S. epicenter at an early stage. We evaluated four metrics related to inter-county trips for NYC: total inflow trips, total outflow trips, the number of origin counties accounting for those inflow trips, and the number of destination counties accounting for those outflow trips (Extended Fig. 1). We found that NYC stayed in the top three even after national emergency in all those four aspects when compared with other counties. This raised concerns on inter-county disease transmissions.

**Influence of the inter-county trips from New York City on the local outbreak**

From January 1 to May 15, the inter-county trips from NYC directly distributed to 42% counties in the United States, including Alaska and Hawaii, and might have a wider impact considering the final destinations of those trips. Despite that the inter-county trips generated from NYC have been reduced by 60% at the most, the overall volume is still high compared with other counties. We first evaluated the influence of such trips on local outbreaks by zooming in on the top twelve destination counties (Fig. 2). Fig. 2b shows the temporal changes of inter-county trips coming from NYC to the top twelve destination counties, and Fig. 2a shows the temporal changes in the cumulative cases in those counties. Nassau County and Westchester County in New York are the two counties with most inter-county trips from NYC before the pandemic. Therefore, they also have the most cumulative cases during the first three weeks after the pandemic began. As the confirmed cases increased, Nassau County and Westchester County experienced a sharp decrease in the trip volume from NYC. In the meantime, the decrease in trips to Suffolk County, NY was not as significant, which later made Suffolk County succeed Nassau County as the county with the second-most confirmed cases. Although Hudson ranked third before pandemic, it has experienced the maximum percentage reduction in the inter-county trips from NYC and thus had a lower rank in confirmed cases.

As the influence of inter-county trips from NYC was unveiled, we further investigated the impacts by examining the day-by-day correlations between such trips and the cumulative COVID-19 cases per capita (per thousand people) of the 1360 destination counties. Since COVID-19 has an incubation period, we have considered four scenarios: no time lag, one-week lag, two-week lag, and three-week lag (Fig. 3). For example, the one-week lag scenario calculates the correlation between the inter-county trip volume from NYC and the cumulative case number per capita one week later. Overall, there is a significant positive correlation between inter-county trips from NYC and the outbreak severity in the destination county. The correlation became more significant as the outbreak kept spreading. And it can be as high as 0.68 (Pearson's r) on Apr 23 in the three-week lag scenario and 0.72 (Spearman's rs) on Apr 22 in the three-week lag scenario. Among the four scenarios, the ones with time lag show a stronger correlation than that without time lag, and the correlation strength is similar between the ones with time lag. We thus conclude that travel from severely infected areas would significantly contribute to local outbreaks with time lags.

**Quantifying the influence of the inter-county trips via "log-linear double-risk" models**

Following the case study in NYC, we expanded our study to quantifying the general influence of inter-county trips from all the infected counties. We innovatively developed a "log-linear double-risk" model, which considers both the external risks of the inter-county trips from infected counties and the internal risks in terms of the vulnerability of local populations. For a given county, the external risk (ER) is measured by the sum of inter-county trips from different counties with each sub-flow weighted by the cumulative cases per capita (per thousand people) in the origin county. The internal risk is defined as a weighted sum of socio-demographic indicators based on high-risk phenotypes[13-15], including percentage of people 65 years and over, percentage of male, percentage of African-American, and median household income. The model is specified as follows.

$$\log(S_j^T) = \alpha \log ER_j + IR_j + \gamma$$

$$ER_j = \sum_{i=1}^{N_j} w_i E_{ij}, w_i = \frac{Cumulative\ Case\ Number_i}{Population_i} \times 1000$$

$$IR_j = \beta_1 Age_j + \beta_2 Male_j + \beta_3 Afri_j + \beta_4 Inc_j$$

Where $S_j^T$ is the outbreak severity of county $j$ measured by the cumulative cases per 1000 people with time lag $T$, where $T$ equals to 0, 1, 2, 3 for no time lag, one-week lag, two-week lag, and three-week lag, respectively; $ER_j$ is the external risk for county $j$; $N_j$ is the number of origin counties that generate inter-county trips to county $j$; $E_{ij}$ is the trips from county $i$ to county $j$, where $i \neq j$; $w_i$ is the weight for inter-county trips originated from county $i$; $IR_j$ is the internal risk for county $j$; $\alpha$ is the coefficient for the logarithm to the base 10 of the external risk; $\beta_1$, $\beta_2$, $\beta_3$, and $\beta_4$ are the coefficients for the internal risks, including the percentage of senior people ($Age_j$), male ($Male_j$), African-American ($Afri_j$), and the county-level median income ($Inc_j$); $\gamma$ is the constant in the model.

We fit the model into twenty scenarios as shown in Fig. 4. The results show that the model with one-week lag outperforms the others when applied to the whole after-pandemic announcement period (the combination of behavior change, quarantine fatigue, and partial reopening) with $R^2$=0.62. For individual stages, the models with shorter time lags have a better fit, which may imply that the community transmissions contribute more to the outbreak severity than the imported cases from other counties. Another possible reason is that most inter-county trips are conducted by residents regularly, e.g., daily, which reduces the significance of an incubation period. Besides, the models fit the data better in the later stages. It may indicate that our assumption for the external risks, where the infected ratio of inter-county travelers is the same as that of the entire population in the origin county, became more literal along with the spread of COVID-19. Moreover, the model shows that the logged external risk is much more important than internal risk by improving R2 significantly (Extended Table 1).

**Discussion**

This study focuses on the population flow between all counties in the U.S. and inspects how inter-county trips aggravate COVID-19 to provide insights into the epidemiological situation in the U.S. First, the study is based on aggregate population flow at a large scale by leveraging integrated mobility location data from over 100 million anonymous devices. Such derived population flow data is quite valuable to help fight against the pandemic. Moreover, our study distinguishes itself for the following reasons. Current studies usually analyze the change of common mobility metrics under the impact of COVID-19, such as percentage of people staying home, miles traveled per person, or number of trips per person[21, 22]. There is a lack of research on how those mobility metrics influence the spread of COVID-19. Very limited research have studied the impact of population outflow from the Chinese epicenter, Wuhan, on the pandemic[5, 23]. However, the epidemiological situation in the U.S. is quite different from China because of multiple epicenters in the early stage, relatively less rigorous travel restrictions, and partial reopening amid COVID-19. This epidemiological situation in the U.S. necessitates an analysis of large-scale population flow between local regions. Our study fills this research gap by analyzing the trips between counties in the U.S., especially the aggravating impact on COVID-19. Not only do we analyze the outflow trips generated by New York City, the largest epicenter in the U.S., we also examine the mobility pattern of the large-scale inter-county trips. We found that the reduction of national total inter-county trips is only 35% at the most. Within four weeks of the national emergency, the inter-county trips begin to rebound nationwide despite ongoing stay-at-home orders, which keep rising. With the partial reopening of society, quantifying the influence of inter-county population flow on the pandemic cannot be ignored. The informative findings regarding spatiotemporal patterns of the inter-county trips in our paper enhance the situational awareness of mobility in the U.S. Furthermore, our paper analyzes the deteriorating effect of inter-county trips on the pandemic using an innovatively-developed "log-linear double-risk" model, which considers both external risk (sum of inflow inter-county trips weighted by local outbreak severity in each trip origin) and internal risk (the vulnerability based on high-risk phenotypes including people age 65 and older, male, African-American, and low-income). Twenty scenarios considering different time lags and periods are compared. As for the whole after-pandemic announcement period, the one-week-time-lagged model had the best fitness (R2=0.62), which is consistent with the research findings that the mean/median incubation period ranges from 4 to 7 days and most cases develop symptoms within 14 days[3, 24-27]. Our study contributes to the non-pharmacological interventions on COVID-19 in the U.S. by analyzing the between-county population flow at a large scale. Especially now with partial reopening started in all 50 states, our model is promising to assist policymakers.

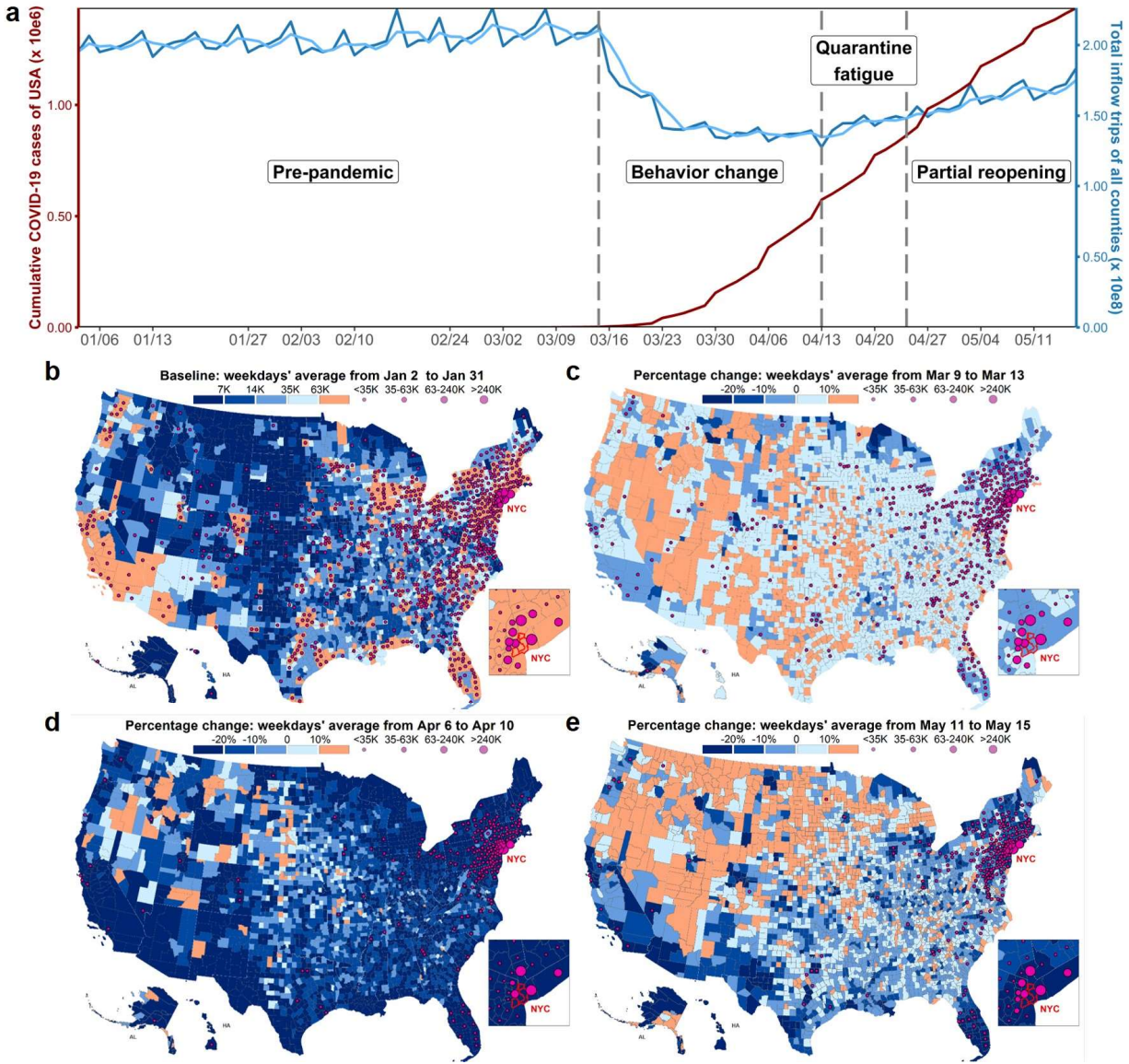

**Fig. 1: Temporal dynamics of national total inflow inter-county trips and spatial differences between counties.** (**a**) The temporal change of national total inflow inter-county trips (dark-blue line) from Jan 2, 2020 to May 15, 2020. Only weekdays and no holidays are plotted with dates on x-axis that are Mondays. The three-day moving average (light-blue line) shows the trend. Meanwhile, national total cumulative COVID-19 cases are shown for comparison (red line). The three dashed dividing lines on dates Mar 13, Apr 13, and Apr 24, respectively, form four stages: pre-pandemic, behavior change, quarantine fatigue, and partial

reopening. (**b**) The blue-to-orange color indicates the baseline of each county: the weekdays' (no holidays) average of inflow inter-county trips by counties from Jan 2 to Jan 31. All counties differ from each other regarding baseline value. The red dots present the spatial distribution of outflow trips produced by New York City and the dot size for each destination county indicates the volume of inflow trips originating in NYC. Considering the overlapped red dots, the NYC area is enlarged in the square frame. (**c**), (**d**), (**e**) The percentage change of inflow inter-county trips by counties relative to the baseline (**b**): the weekdays' (no holidays) average for the week of national emergency (**c**), the week with the least inflow trips (**d**), and the most recent week of studied period (**e**). The red dots have the same meaning as (**b**). Although long-distance trips from NYC clearly decrease after the national emergency is announced, NYC keeps influencing nearby counties.

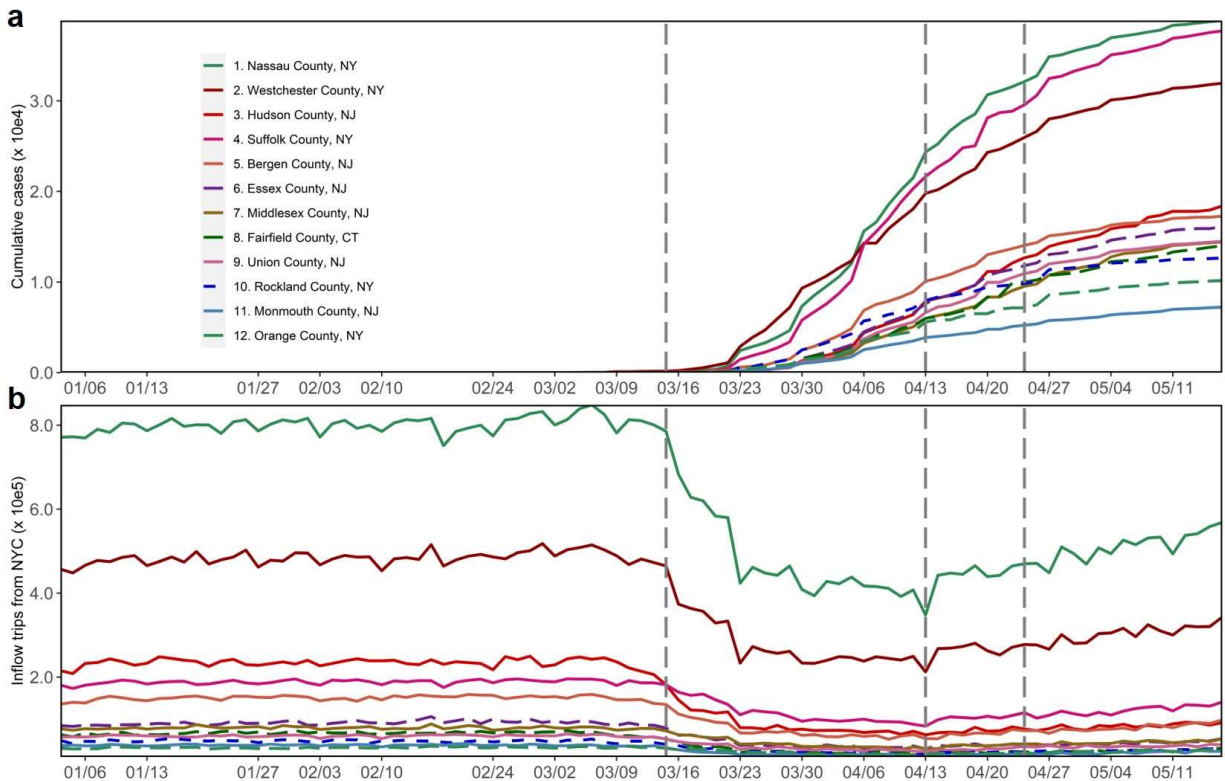

**Fig. 2: Top twelve destination counties with the most inflow trips coming from New York City. (a)** The temporal changes of the cumulative cases by the twelve destination counties. The number for each legend label indicates the rank in terms of inflow from NYC. For instance, Nassau is the top county receiving the most trips from NYC. **(b)** The temporal changes of inflow trips from NYC of those twelve counties. Each destination county shares a similar pattern with the national trend in Fig. 1a. The three dividing lines show the four stages and dates on x-axis are Mondays during the studied period (Jan 2 to May 15, only weekdays and no holidays)

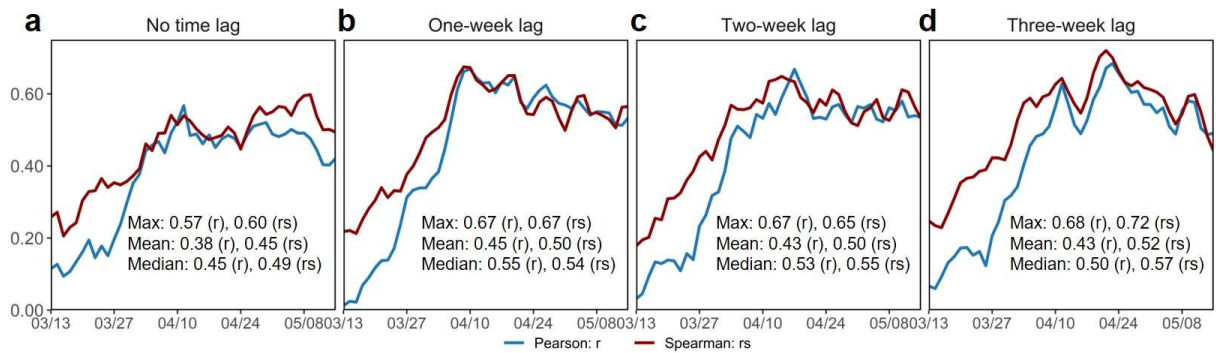

**Fig. 3: Daily correlation coefficients between inflow trips from NYC and cumulative cases per capita when applied to all destination counties receiving trips from NYC.** Different time lags between the inflow trips and the cumulative cases per capita are compared: no time lag (**a**), one-week lag (**b**), two-week lag (**c**), and three-week lag (**d**). The dates on x-axis correspond to the cumulative cases per capita, which range from Mar 13 to May 15 (only weekdays, no holidays). All three subplots show both Pearson and Spearman coefficients. In addition, each subplot presents the maximum, mean, and median of Pearson and Spearman coefficients during the studied period.

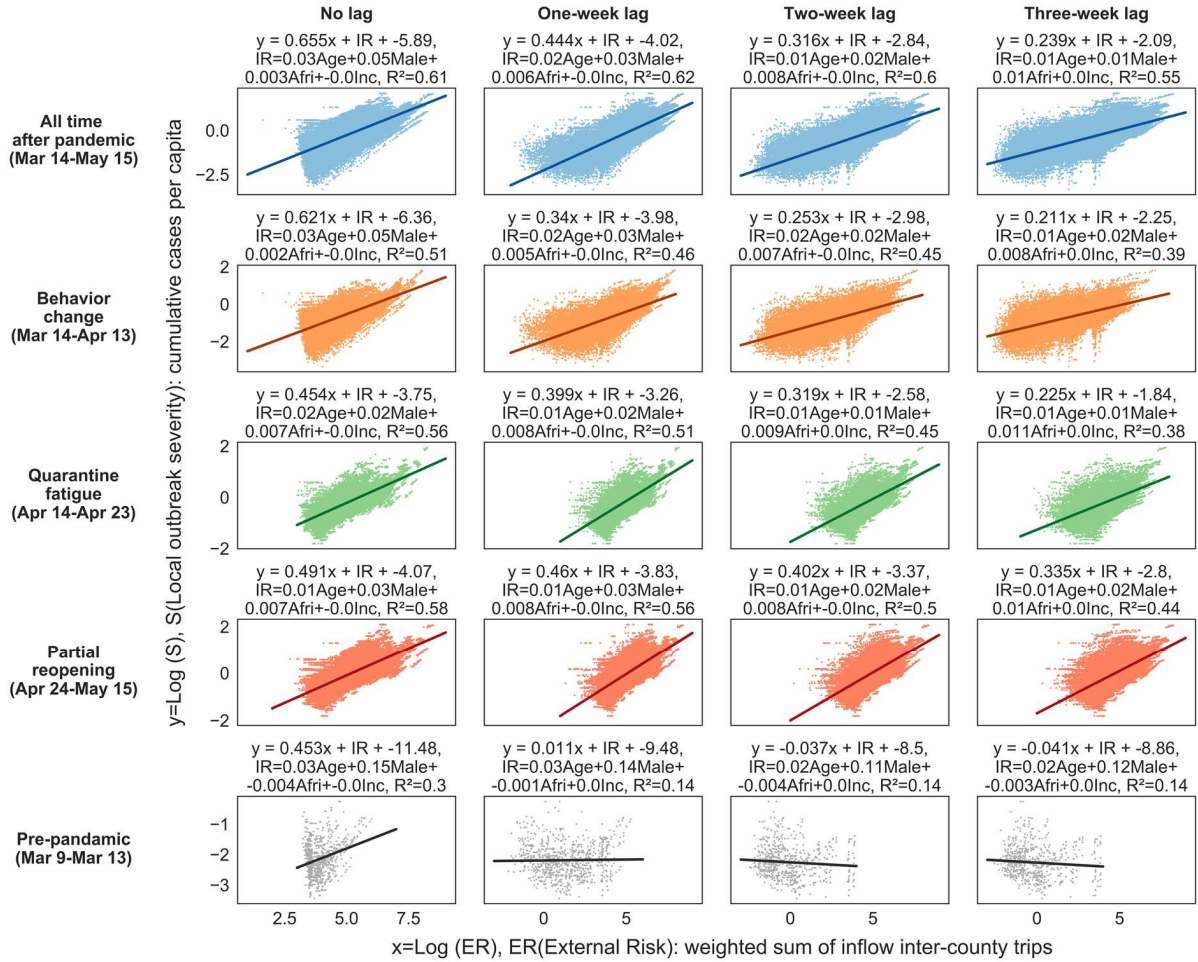

**Fig. 4: Modeling results under twenty scenarios with different periods and time lags and the corresponding scatter plots between logged external risk and logged local outbreak severity.** Five periods (all time after pandemic begins, behavior change, quarantine fatigue, partial reopening, and pre-pandemic from top to bottom) and four types of time lag (no lag, one-week, two-week, and three-week lag from left to right) comprise the twenty scenarios. Each scatter plot shows the association of logged external risk with the logged local outbreak severity along with a fitted regression line. In addition, the title for each subplot presents the complete form of estimated model and corresponding $R^2$. The developed model is more appropriate to the after-pandemic period. In addition, no time lag and one-week lag work better than the others.

## Acknowledgments

We would like to thank and acknowledge our partners and data sources in this effort: (1) partial financial support from the U.S. Department of Transportation's Bureau of Transportation Statistics and the National Science Foundation's RAPID Program; (2) Amazon Web Service and its Senior Solutions Architect, Jianjun Xu, for providing cloud computing and technical support; (3) computational algorithms developed and validated in a previous USDOT Federal Highway Administration's Exploratory Advanced Research Program project; and (4) COVID-19 confirmed case data from the Johns Hopkins University Github repository and sociodemographic data from the U.S. Census Bureau.

## Author Contributions

LZ and QS designed the study. QS, YP, and WZ analyzed the data. QS, YP, and WZ interpreted the data. QS, YP, and WZ wrote the manuscript. All authors contributed to the final draft.

## Competing Interests

All authors declare no competing interests.
## Additional Information

The aggregated inter-county trips will be published on a public site at data.covid.umd.edu. The authors declare no competing financial interests. Correspondence and requests for materials should be addressed to QS (qsun12@umd.edu). Reprints and permissions information is available at www.nature.com/reprints.

## Extended Data

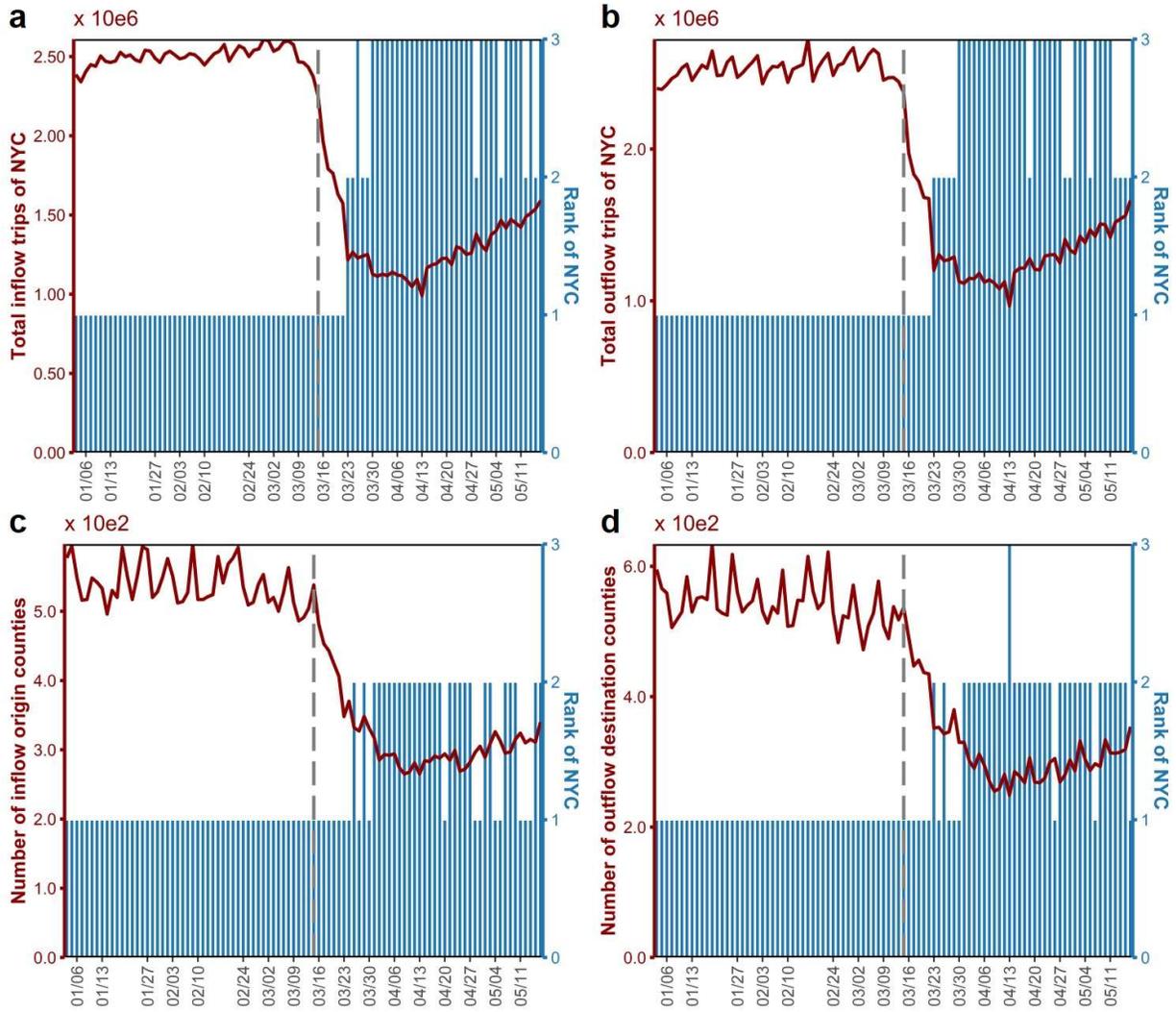

**Fig. 1: The daily inflow and outflow trips of New York City and the corresponding number of origin and destination counties along with the ranking. (a), (b),** The daily total inflow trips (**a**) and total outflow trips (**b**) of NYC, respectively, from Jan 2 to May 15 (only weekdays, no holidays) with x-axis showing the included Mondays. (**c**), The daily number of origin counties accounting for those inflow trips coming to NYC. (**d**), The daily number of destination counties accounting for those outflow trips produced by NYC. The blue bars for all subplots show the daily corresponding rank of NYC when comparing it with other counties. The rank is in

descending sequence; for example, NYC stays in the top three with the most inflow trips (**a**). The dividing line on Mar 13 is used for comparing the pre- and post- pandemic situation of NYC.

**Table 1: Importance measure of logged external risk under sixteen scenarios.** This table shows the importance of logged external risk relative to the internal risk for modeling result (R2) under scenarios of different periods and types of time lag. For each scenario, the model is built twice: with and without permuting logged external risk. Permuting logged external risk means that the model only includes internal risk, while not permuting means both logged external risk and internal risk are included in the model. For any scenario, the logged external risk shows its significant importance by improving R2 value.

| Periods | Permuting ER or not | No lag | One-week lag | Two-week lag | Three-week lag |
|---|---|---|---|---|---|
| All time after pandemic | Yes | 0.06 | 0.06 | 0.07 | 0.08 |
| | No | 0.61 (+0.55) | 0.62 (+0.56) | 0.60 (+0.53) | 0.55 (+0.47) |
| Behavior change | Yes | 0.04 | 0.04 | 0.04 | 0.05 |
| | No | 0.51 (+0.47) | 0.46 (+0.42) | 0.45 (+0.41) | 0.39 (+0.34) |
| Quarantine fatigue | Yes | 0.23 | 0.23 | 0.23 | 0.23 |
| | No | 0.56 (+0.33) | 0.51 (+0.28) | 0.45 (+0.22) | 0.38 (+0.15) |
| Partial reopening | Yes | 0.23 | 0.23 | 0.23 | 0.23 |
| | No | 0.58 (+0.35) | 0.56 (+0.33) | 0.50 (+0.27) | 0.44 (+0.21) |